\documentclass[gmd, manuscript]{copernicus}
\usepackage{amsmath}
\usepackage{array}

\begin{document}

\title{An exact finite-state representation of viscoelastic memory for accelerated sea-level modeling: SLEvolve v1.0}

\Author[1][xiyuanbao@fas.harvard.edu]{Xiyuan}{Bao}

\Author[1]{S. Mostafa}{Mousavi}

\Author[2]{Konstantin}{Latychev}

\Author[1]{Jerry X.}{Mitrovica}

\affil[1]{Department of Earth and Planetary Sciences, Harvard University, Cambridge, MA 02138, USA}

\affil[2]{Seakon, Toronto, ON, Canada}

\runningtitle{Exact finite-state viscoelastic memory: SLEvolve v1.0}
\runningauthor{Bao et al.}

\received{}
\pubdiscuss{}
\revised{}
\accepted{}
\published{}

\firstpage{1}

\maketitle

\nolinenumbers
\linespread{1.0}\selectfont

\begin{abstract}
Glacial Isostatic Adjustment (GIA) models solve the gravitationally self-consistent sea-level equation for the response of the solid Earth and the oceans to changing ice loads.
As these models are increasingly embedded in coupled ice-sheet simulations, glacial-cycle calculations, and large ensembles, standard pseudo-spectral 1-D solvers face a steep computational barrier: they evaluate the viscoelastic response by explicitly re-summing every past load increment, scaling quadratically with time steps as $O(N^2)$.
Here we show that this computational penalty is an artifact of algorithmic formulation rather than Earth physics. By exploiting the exponential form of the viscoelastic relaxation kernel, we reformulate the viscous response as an exact recursive update of a finite set of modal states, without re-summing the load history.
This exact reformulation alone reduces the cost of a run to $O(N)$, yielding a 20.6-fold speed-up relative to a standard full-history 1-D GIA solver for a case study involving a 550-step Antarctic future sea-level projection. Implemented in our open-source solver SLEvolve v1.0, this formulation---combined with fewer, table-free and multithreaded spherical-harmonic transforms, improved file handling, and physics-inspired iteration starts---achieves an end-to-end speed-up of over three orders of magnitude (up to 4,000-fold) against the same 1-D GIA solver, reducing day-long simulations to under thirty seconds on a workstation. Beyond this solid-Earth application, SLEvolve illustrates how human-derived analytical reformulations can be combined with benchmark-harnessed AI-assisted engineering to modernize scientific software. In particular, Numerical regression tests and performance profiling guided the AI agents' implementation choices and helped reject ineffective optimizations.

\end{abstract}

\introduction

Sea-level change at any given time driven by the long term (> Maxwell time) redistribution of ice and water depends on the entire preceding loading history because past load changes continue to contribute over time scales governed by the Earth’s spectrum of viscoelastic relaxation times (Fig.~\ref{fig:concept}a). Models that solve the gravitationally self-consistent sea-level equation \citep{farrell1976,kendall2005} must therefore account for this loading history throughout the calculation. Such models are increasingly used at fine temporal resolution in century- to millennial-scale projections and ice-sheet coupling \citep{gomez2010,han2022}, in large ensemble calculations \citep{caron2018,lin2021}, and repeatedly within the glacial-cycle simulations that iteratively solve for the initial bedrock topography \citep{kendall2005,gomez2013}. In all of these applications, the computational cost of evaluating the loading-history contribution can become a principal limitation on the adopted spatial and temporal resolution or on the number of simulations that are practical.

The widely adopted pseudo-spectral approach to solving the sea-level equation \citep{mitrovica1991,kendall2005} is based on viscoelastic normal-mode theory valid for a 1-D (radially varying) Earth structure \citep{peltier1974,tromp2000}. In normal mode theory, the viscoelastic impulse response of the Earth model is first obtained in the Laplace domain and then transformed into time-dependent Love numbers that are functions of the spherical harmonic degree. These numbers are combined to derive a sea level Green's function, and the sea level response can be computed at each degree via a convolution of these functions with the loading history.  The latter is generally expressed as the sum of a series of past load increments.  \citet{kendall2005} integrated the viscoelastic response to each discrete load increment analytically in time and at step $j$ all earlier increments are revisited, leading to a total cost that scales as $O(N^2)$ for $N$ time steps (Fig.~\ref{fig:concept}b). This full-history formulation is adopted in all the open sea-level codes that we have examined, including SLcode \citep{austermann2015}, SELEN4 \citep{spada2019}, giapy \citep{kachuck2019}, and the solver of \citet{han2022}, which builds on the code of \citet{goldberg2016}. We denote the ``standard simulation'' code of \citet{han2022} without their ``time window'' as the legacy solver and use it below as the full-history reference. Our results are equally applicable to the other solvers listed above.

Previous efforts to reduce the cost of 1-D GIA solvers primarily focus on limiting the number of past load increments in the sum, either by truncating the memory of each increment or by coarsening older steps. \citet{deboer2014} compute the response to each new load increment once at 15 prescribed future times over an 80-kyr interval and accumulate these responses in a stored future sea-level field. Their cost per time step is independent of the number of preceding steps, while the response of each increment is truncated after 80 kyr and interpolated between the prescribed times. \citet{han2022} instead propose a time-window scheme that combines older loads at progressively coarser temporal resolution. They report that this changes the growth of total CPU time from quadratic to linear and halves the runtime of the 550-step future projection and a 1200-step run with two glacial cycles. The reduced load history in the time-window version of the code in \citet{han2022} introduces an very small error, about 1\,m or 0.2\,\% of the deformation for the 550-step case, relative to the full-history solution (Sect.~\ref{sec:exactness}).

The schemes above reduce the cost of the history term by changing when, or at what temporal resolution, the contribution of each load increment is retained. An alternative is to carry the accumulated viscoelastic memory forward as an evolving state. This idea has a long history outside 1-D sea-level formulations. In computational viscoelasticity, relaxation kernels represented by sums of exponentials are evaluated recursively using internal variables, one for each exponential term \citep{taylor1970,simo1998}.
Within GIA, \citet{wu1982} noted that, for a load whose assumed prehistory has the same time dependence globally, that prehistory enters the response only through one amplitude per relaxation mode, which they evaluated once to correct for initial isostatic disequilibrium.

State-based descriptions have also emerged in several reduced GIA models. \citet{bueler2007} showed that the viscous half-space component of the elastic-plate--viscous-half-space model of \citet{lingle1985}, later termed ELVA and originally evaluated by convolution with the full load history, can be written as a time-dependent equation and advanced as a displacement state in Fourier space, with one state variable for each wavenumber. A simpler reduced model, ELRA, represents the Earth response by a single prescribed relaxation time and advances displacement toward its instantaneous equilibrium \citep{lemeur1996}. FastIsostasy extends this state-based reduced-model approach to regional calculations in which laterally variable lithospheric thickness and viscosity are represented approximately, by scaling the equation of an elastic plate on a single effective viscous half-space column by column \citep{swierczek2024}; the depth-dependent, multi-mode viscoelastic response and the global sea-level equation are not solved in that approach. Three-dimensional time-domain GIA codes such as VILMA \citep{martinec2000,hagedoorn2007,klemann2008}, CitcomSVE \citep{zhong2003,zhong2022,yuan2025} and SeaKon \citep{latychev2005,gomez2018,lau2026} likewise carry viscoelastic memory in the evolving mechanical state of the Earth, advancing stress and deformation from one time step to the next rather than replaying the complete loading history; however, they differ in how the Maxwell law is stepped in time (Table~\ref{tab:lineage}).

In the standard 1-D pseudo-spectral, self-gravitating sea-level formulation \citep{kendall2005,gomez2010,austermann2015,spada2019,kachuck2019,han2022} the response to each step-wise load increment is evaluated from precomputed Love numbers analytically in time when these are in normal-mode form and from numerically tabulated time series in giapy \citep{kachuck2019}, but the contributions from all preceding increments are still explicitly re-summed at every step. In the normal-mode representation, or its collocation form in which the decay rates are prescribed on a logarithmic grid and the residues fitted \citep{mitrovica1992}, however, this viscous response is represented by a finite sum of exponentially decaying terms \citep{peltier1985,mitrovica1992}. Each exponential contribution can therefore be mapped directly to an evolving modal state. The present work applies this construction to the full 1-D gravitationally self-consistent sea-level equation, eliminating the explicit history-summation bottleneck without truncating or coarsening the prescribed loading history. This cost is not a limitation of the pseudo-spectral formulation itself \citep{mitrovica1991,milne1998,kendall2005}. In our view, the pseudo-spectral solution of the sea-level equation, with its inner iteration for the ocean load and outer iteration for the initial topography, remains the most efficient way to solve the 1-D problem. We retain these components unchanged and modify only the evaluation of the viscous history term. Unlike time-domain GIA codes which evolve spatially distributed viscoelastic states, such as local stress, the modal states introduced here directly accumulate the global relaxation response to surface loading through the precomputed Love-number spectrum. For the prescribed step-wise loading, our exponential updates reproduce the full-history response without additional time-integration error. Table~\ref{tab:lineage} summarizes these parallel developments and the position of the present formulation among them.

\begin{table*}[!tp]
\linespread{1.0}\selectfont
\caption{How GIA formulations and related viscoelastic methods carry past loading forward in time, grouped by formulation family. Bold marks the physics represented; the last row (this work) is set in bold throughout. The last column lists approximations beyond the step-wise load history common to all formulations. SLE: sea-level equation.}
\label{tab:lineage}
\scriptsize\setlength{\tabcolsep}{3pt}\renewcommand{\arraystretch}{1.0}
\begin{tabular}{@{}>{\raggedright\arraybackslash}p{2.2cm}>{\raggedright\arraybackslash}p{3.45cm}>{\raggedright\arraybackslash}p{3.3cm}>{\raggedright\arraybackslash}p{2.75cm}>{\raggedright\arraybackslash}p{1.6cm}>{\raggedright\arraybackslash}p{3.3cm}@{}}
\tophline
Formulation & Representative work & Earth response represented & How past loading is carried & Cost of a run of $N$ steps & What is approximated\\
\middlehline
General viscoelasticity: recursive convolution for material laws & finite-element viscoelasticity \citep{taylor1970,simo1998} & \textbf{not an Earth model:} one material point, relaxation function fitted as a sum of exponentials & state: one internal variable per exponential term, updated each step & $O(N)$ & strain linear in time over each step, or step integral by the midpoint rule\\
\middlehline
Reduced Earth: ELVA (elastic plate, viscous half-space), state form & \citet{bueler2007} & \textbf{flat, uniform viscous half-space under an elastic plate (the ELVA model of \citealp{lingle1985}); wavelength-dependent relaxation;} elastic part: spherical, self-gravitating Green's function \citep{farrell1972}; no sea-level equation & state: displacement advanced in Fourier space, one mode per wavenumber & $O(N)$ & trapezoid (Crank--Nicolson) step; periodic FFT (Fast Fourier Transform) domain; elastic part by convolution\\\noalign{\vskip 2.5pt}
Reduced Earth: ELRA (elastic lithosphere, relaxing asthenosphere) & \citet{lemeur1996} & \textbf{flat elastic plate; one relaxation time for all wavelengths;} no self-gravitation or sea-level equation & state: displacement relaxed toward equilibrium & $O(N)$ & relaxation equation integrated with finite time steps\\\noalign{\vskip 2.5pt}
Reduced Earth: regional, laterally variable & FastIsostasy \citep{swierczek2024} & \textbf{flat, regional; laterally variable plate thickness and viscosity;} no self-gravitation; regional sea-surface change only; no global sea-level equation or rotation & state: displacement advanced by a differential equation & $O(N)$ & depth structure lumped into one effective viscous half-space; lateral variation by column-wise scaling of the plate equation, not a 3-D flow solution; adaptive explicit time steps\\
\middlehline
3-D viscoelastic Earth, time domain & VILMA \citep{martinec2000,hagedoorn2007,klemann2008}; CitcomSVE \citep{zhong2003,zhong2022,yuan2025}; SeaKon \citep{latychev2005,gomez2018,lau2026} & \textbf{global, self-gravitating; 3-D viscosity and lithospheric thickness;} sea-level equation, shoreline migration and rotation only in some versions & state: current stress (viscous memory) updated each step & $O(N)$, with a full 3-D solve per step & Maxwell law time-stepped (explicit, first order: VILMA; trapezoid, second order: CitcomSVE; exponential propagator, exact for constant strain rate, iterated within the step: SeaKon); spatial discretization\\
\middlehline
1-D normal-mode pseudo-spectral sea-level equation, full history & formulation \citep{kendall2005}; SLcode \citep{austermann2015}; SL\_SLG\_TPW code of \citet{goldberg2016}, partly adopted by \citet{han2022};  SELEN4 \citep{spada2019}; giapy \citep{kachuck2019} & \textbf{global, self-gravitating; sea-level equation} with migrating shorelines and rotation; \textbf{layered (1-D) viscosity}; viscous response a sum of relaxation modes (giapy: Love numbers tabulated in time) & sum over all past load increments, re-evaluated each step & $O(N^2)$: one term per past step & none: response to each step-wise load increment is analytic in time (reference formulation)\\\noalign{\vskip 2.5pt}
1-D normal-mode SLE, precomputed future response & \citet{deboer2014} & as 1-D full history & future response of each increment computed once, added to a stored future sea-level array & $O(N)$, fixed set of future times & response truncated at 80\,kyr and interpolated from 15 times; shoreline feedback only approximate\\\noalign{\vskip 2.5pt}
1-D normal-mode SLE, time window & \citet{han2022} & as 1-D full history & as full history, older increments merged into coarser steps & $O(N)$ (reported) & older load states sub-sampled at coarser steps (increments merged)\\
\middlehline
\textbf{This work: recursive normal-mode state} & \textbf{SLEvolve v1.0} & \textbf{as 1-D full history} & \textbf{state: one field per relaxation mode, updated each step; complete history kept} & \textbf{$O(N)$ from the first step: one update per mode} & \textbf{none: exact exponential update, identical to the full-history sum}\\
\bottomhline
\end{tabular}
\end{table*}

This paper makes five contributions.
\begin{itemize}
\item[(i)] We establish an exact recursive state-space representation for variable time steps and all six viscous history terms of the normal-mode sea-level equation, reducing the history-summation cost from $O(N^2)$ to $O(N)$. A state-update placement rule preserves equivalence to the full-history response within the nonlinear sea-level iteration (Sect.~\ref{sec:methods}).
\item[(ii)] We further reduce the end-to-end cost through implementation measures, including a state kept in memory across steps and binary input, removal of redundant transforms, faster spectral transforms and multi-threading (Sect.~\ref{sec:implementation}).
\item[(iii)] We introduce three workflow measures that reduce the number of iterations: an extrapolated initial guess for the inner iteration at each time step and, for ice age calculations, a partially relaxed initial-topography guess and a warm start of the inner iteration from the previous outer pass (Sects.~\ref{sec:extrap}, \ref{sec:paleo} and \ref{sec:iceage}).
\item[(iv)] We verify SLEvolve against the legacy solver through machine-precision regression benchmarks in one forward projection and two ice age scenarios, documenting agreement below $10^{-11}$\,m and end-to-end speed-ups up to 4000-fold (Sects.~\ref{sec:verification}--\ref{sec:results}).
\item[(v)] We release SLEvolve v1.0 with this paper in both Fortran and Python, providing independent implementations of the recursive formulation and the benchmark cases used here.
\end{itemize}

Beyond its geophysical utility, SLEvolve provides a case study in scientific software modernization that combines human-derived mathematical reformulation with benchmark-harnessed AI coding agents. Authors established the analytical formulation and solver design, while the agents explored implementation and workflow alternatives under numerical regression tests and runtime profiling. These benchmarks also helped identify and reject approaches whose overhead outweighed their benefits (Sect.~\ref{sec:ai}; Sect.~\ref{sec:discussion}).

\section{Methods}
\label{sec:methods}

\subsection{Sea-level equation and normal-mode Love numbers}
\label{sec:sle}

We follow the pseudo-spectral formulation of \citet{kendall2005} as implemented by \citet{han2022}.
The time steps are $t_0<t_1<\dots<t_N$ with intervals $\delta t_j=t_j-t_{j-1}$, which can be non-uniform.
At each step the solver finds the change in ocean height since $t_0$, $\Delta S_j$, such that it is gravitationally self-consistent with the Earth's response to the ice and ocean loads and with conservation of surface mass. Sea level is defined here as the vertical distance between the sea surface and the solid Earth surface. The geometry is described by the ocean function $C_j$ (1 where the sea surface is above the solid Earth, i.e., sea level is positive, and 0 elsewhere), the ice function $\beta_j$ (0 where grounded ice covers the surface, 1 elsewhere; marine ice counts as grounded where its thickness exceeds the water depth times $\rho_w/\rho_I$, following \citealp{kendall2005}).

This is the generalized sea-level equation of \citet[][their Eqs.~9--11 and 17]{kendall2005},
\begin{equation}
\Delta S_j = \Delta\mathcal{SL}_j\,C_j\beta_j - T_0\,(C_j\beta_j-C_0\beta_0),\qquad \Delta\mathcal{SL}_j=\Delta\mathcal{G}_j-\Delta\mathcal{R}_j+\frac{\Delta\Phi_j}{g},
\label{eq:sle}
\end{equation}
where $\Delta\mathcal{R}$ is the change in the radial position of the solid surface, $\Delta\mathcal{G}$ is the geographically variable component of the change in the sea-surface equipotential (or, equivalently, the geoid change), $g$ is the surface gravity, $\Delta\Phi_j/g$ is the uniform shift of the sea-surface equipotential computed by invoking the conservation of ice-plus-ocean mass \citep[][their Eq.~36]{kendall2005} (the uniform term below), and $T_0$ the topography at $t_0$, which is prescribed in a forward run and unknown in an ice age run.
Following \citet{kendall2005}, $\Delta$ denotes changes since time $t_0$, while $\delta$ denotes increments between successive time steps. Eq.~(\ref{eq:sle}) is nonlinear because the ocean function $C_j$ and, through the flotation check, the ice function $\beta_j$ depend on the sea level being solved for.

Following \citet{kendall2005}, the algorithm for solving Eq.~(\ref{eq:sle}) involves two layers of iterations. At every time step, after the next ice load increment is applied, a Picard (fixed-point) iteration on the ocean-load increment $\delta S_j$ is performed to convergence; this is the so-called inner iteration. In practice, this iteration stops when the relative change of $\delta S_j$ between successive iterates falls below some prescribed tolerance $\epsilon$. We adopt $\epsilon=10^{-5}$ in the calculations below, which is stricter than the value of $\sim10^{-4}$ suggested by \citet{kendall2005}, so that the comparison between solvers is not limited by the tolerance. In ice age calculations, a second, ``outer iteration'' is performed in which the full, start-to-finish, ice history is repeated to successively revise the initial topography, $T_0$, such that the system evolves to a present-day topography field within some prescribed tolerance of the observed present topography, $T_{\mathrm{obs}}$.

For every spherical-harmonic coefficient $(\ell,m)$ the sea-level response to the load decomposes into an elastic and a viscous part; for brevity we rewrite Eq.~B17 of \citet{kendall2005} for a non-rotating Earth in compact form,
\begin{equation}
\Delta\mathcal{SL}_{\ell m}(t_j)=T_\ell\left[E_\ell\,\Delta L_{\ell m}(t_j)+V_{\ell m}(t_j)\right],
\label{eq:responseA}
\end{equation}
\noindent where
\begin{equation}
\quad T_\ell=\frac{4\pi a^3}{M_e(2\ell+1)} ~{\rm and} \quad E_\ell=1+k^E_\ell-h^E_\ell,
\label{eq:response}
\end{equation}
with $a$ and $M_e$ the radius and mass of the Earth, the load change $\Delta L_{\ell m}(t_j)=\sum_{n=0}^{j}\delta L_{\ell m}^n$ and the load increment $\delta L_{\ell m}^n=\rho_I\,\delta I^{*\,n}_{\ell m} + \rho_w\,\delta S_{\ell m}^n$, where $\delta I^{*}$ is the increment of the grounded ice thickness, including marine ice where it is grounded, and $\rho_I$ and $\rho_w$ are the densities of ice and water, respectively. The first term in the square brackets is the elastic component of the sea-level response; the second, $V_{\ell m}$, is the viscous or hereditary response, and it is this term that we aim to optimize. The general form of $V_{\ell m}$ is the same for the rotating and the non-rotating Earth. At degree 2 the rotational feedback adds the analogous term $g^{-1}\left[E^{T}_{2}\,\Delta\Lambda_{2m}(t_j)+V^{T}_{2m}(t_j)\right]$, driven by the perturbation of the centrifugal potential $\Lambda$ that follows the load-induced change of the rotation vector, with $E^{T}_\ell=1+k^{T,E}_\ell-h^{T,E}_\ell$ and $V^{T}$ formed as $V$ below but with the tidal Love numbers $k^{T}_\ell$, $h^{T}_\ell$ in place of $k_\ell$, $h_\ell$ \citep[Eqs.~B25--B28]{kendall2005}. These terms are treated below as additional history sums.

The viscoelastic Love numbers are
\begin{equation}
h_\ell(t)=h^E_\ell\,\delta(t)+\sum_{k=1}^{K_\ell} r^\ell_k\,e^{-s_k^\ell t},\qquad k_\ell(t)=k^E_\ell\,\delta(t)+\sum_{k=1}^{K_\ell} r_k^{\ell\prime}\,e^{-s_k^\ell t},
\label{eq:love}
\end{equation}
where $h_\ell$ and $k_\ell$ are the load Love numbers for radial displacement and gravitational potential, respectively, so that $1+k_\ell-h_\ell$ gives the relative sea-level response (sea surface minus solid surface), with the leading 1 representing the direct gravitational-potential contribution of the load; $s_k^\ell>0$ are the inverse relaxation times, and $r_k^\ell$ and $r_k^{\ell\prime}$ are the corresponding modal amplitudes (residues). The symbol $\delta()$ is the Dirac delta function. The Love numbers of the three test cases (Sect.~\ref{sec:cases}) contain $K_\ell=16$ to 26 terms per degree.
These terms correspond either to physical normal modes or to collocation poles prescribed on a logarithmic grid \citep{peltier1985,mitrovica1992}; in both cases the kernel is a finite exponential sum, which is all the recurrence requires.

\begin{figure*}[!tp]
\includegraphics[width=\textwidth]{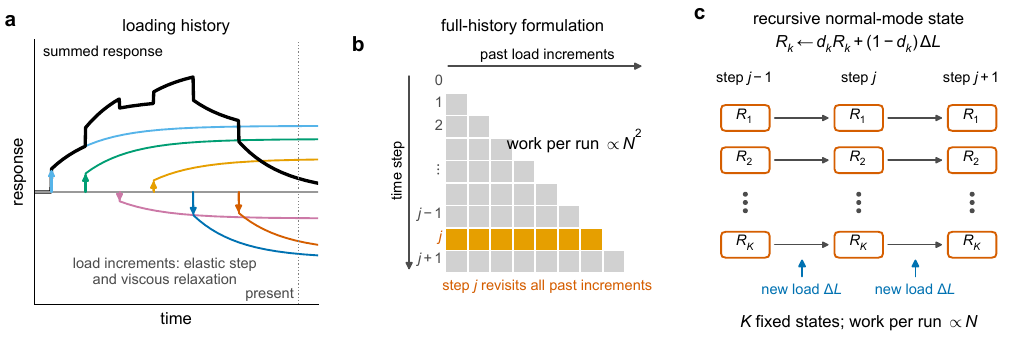}
\caption{Concept.
(a) Load increments and their exponential relaxation sum into the present response.
(b) Full-history formulation: every step revisits all past increments, work $\propto N^2$.
(c) Recursive normal-mode state: a fixed set of $K$ modal states $R_k$ is advanced once per step with the new cumulative load $\Delta L$, work $\propto N$.}
\label{fig:concept}
\end{figure*}

Because each load increment persists after it is applied, its viscous response is the time integral of the impulse response in Eq.~(\ref{eq:love}), $\int_0^{t} r_k^\ell e^{-s_k^\ell\tau}\,d\tau=(r_k^\ell/s_k^\ell)(1-e^{-s_k^\ell t})$ for each mode.
Using these expressions, the hereditary (viscous) part of Eq.~(\ref{eq:responseA}) is the explicit load-history sum
\begin{equation}
V_{\ell m}(t_j)=\sum_{n=0}^{j-1}\beta_\ell(t_j-t_n)\,\delta L_{\ell m}^n,\qquad \beta_\ell(t)=\sum_{k=1}^{K_\ell} a_k^\ell \left(1-e^{-s_k^\ell t}\right),\qquad a_k^\ell =\frac{r_k^{\ell\prime}-r_k^\ell}{s_k^\ell},
\label{eq:history}
\end{equation}
for the sea-level response (the kernel $\beta_\ell$ is Eq.~B14 of \citealp{kendall2005}, not the ice function $\beta_j$). Figure~\ref{fig:concept}a,b illustrates Eq.~(\ref{eq:history}): every step revisits all previous increments.

The full-history reference solver evaluates six such sums. Two are driven by the surface load and provide the sea-surface height and radial-displacement responses; four additional degree-2 sums arise from rotational feedback, namely the inertia perturbation, the rotation vector, and the rotational potential acting on sea surface height and radial displacement. These use the tidal Love numbers for the response to a potential rather than to a surface load \citep{kendall2005}.

\subsection{Recursive normal-mode state}
\label{sec:recursion}

Substituting the kernel $\beta_\ell(t)$ into the expression for $V_{\ell m}(t_j)$ in Eq.~(\ref{eq:history}) and exchanging the order of summation gives
\begin{equation}
V_{\ell m}(t_j)=\sum_{n=0}^{j-1}\sum_{k=1}^{K_\ell} a_k^\ell\left(1-e^{-s_k^\ell(t_j-t_n)}\right)\delta L_{\ell m}^n
=\sum_{k=1}^{K_\ell} a_k^\ell\Big[\sum_{n=0}^{j-1}\delta L_{\ell m}^n-\sum_{n=0}^{j-1}e^{-s_k^\ell(t_j-t_n)}\delta L_{\ell m}^n\Big].
\label{eq:split}
\end{equation}
Where
\begin{equation}
Q_{\ell m}^k(t_j)=\sum_{n=0}^{j-1}e^{-s_k^\ell(t_j-t_n)}\delta L_{\ell m}^n,\qquad
R_{\ell m}^k(t_j)=\Delta L_{\ell m}(t_{j-1})-Q_{\ell m}^k(t_j),
\label{eq:defs}
\end{equation}
so that
\begin{equation}
V_{\ell m}(t_j)=\sum_{k=1}^{K_\ell}a_k^\ell\,R_{\ell m}^k(t_j).
\label{eq:vstate}
\end{equation}
Here $R_{\ell m}^k$ is termed the modal state; it admits a recursive update.
Since $t_j-t_n=\delta t_j+(t_{j-1}-t_n)$, every weight in $Q_{\ell m}^k(t_j)$ carries the common factor $d_k^\ell({\delta t_j})=e^{-s_k^\ell\delta t_j}$, the one-step decay, which can be taken outside the sum:
\begin{equation}
Q_{\ell m}^k(t_j)=d_k^\ell(\delta t_j)\sum_{n=0}^{j-1}e^{-s_k^\ell(t_{j-1}-t_n)}\delta L_{\ell m}^n
=d_k^\ell(\delta t_j)\Big[\underbrace{\sum_{n=0}^{j-2}e^{-s_k^\ell(t_{j-1}-t_n)}\delta L_{\ell m}^n}_{Q_{\ell m}^k(t_{j-1})}+\delta L_{\ell m}^{j-1}\Big].
\label{eq:qrec}
\end{equation}
Since $\Delta L_{\ell m}(t_{j-1})=\Delta L_{\ell m}(t_{j-2})+\delta L_{\ell m}^{j-1}$, substituting this and Eq.~(\ref{eq:qrec}) into the definition of $R$ in Eq.~(\ref{eq:defs}) yields the discrete state-space update:
\begin{equation}
\begin{aligned}
R_{\ell m}^k(t_j)&=\Delta L_{\ell m}(t_{j-1})-d_k^\ell(\delta t_j)\left[\Delta L_{\ell m}(t_{j-2})-R_{\ell m}^k(t_{j-1})+\delta L_{\ell m}^{j-1}\right]\\
&=d_k^\ell(\delta t_j)\,R_{\ell m}^k(t_{j-1})+(1-d_k^\ell(\delta t_j))\,\Delta L_{\ell m}(t_{j-1}),
\end{aligned}
\label{eq:recursion}
\end{equation}
with $R_k(t_0)=0$.

The explicit sum over past increments has therefore been eliminated without approximation, and Eqs.~(\ref{eq:history}) and (\ref{eq:vstate})--(\ref{eq:recursion}) are algebraically identical for any sequence of increments. Each $R_{\ell m}^k$ is a relaxing state driven by the cumulative load $\Delta L_{\ell m}$, with relaxation rate $s_k^\ell$ (Fig.~\ref{fig:concept}c).
A variable step $\delta t_j$ changes only $d_k^\ell$.
The same set of states serves the sea-surface height and radial-displacement responses, which differ only in their modal weights.
The horizontal (tangential) displacement is obtained analogously from the same states using the horizontal-displacement load Love number $l_\ell$: its viscous component has modal weights $r_k^{\ell,l}/s_k^\ell$, while its instantaneous elastic component is determined by $l_\ell^E$.
It is available as an optional output of SLEvolve. The four rotational sums at degree 2 carry their own states with the same update, using the tidal Love numbers for the response to the rotational potential.

\subsection{Complexity and storage}
\label{sec:complexity}

With Eq.~(\ref{eq:history}), step $j$ evaluates $j$ kernel terms, each a sum over $K_\ell$ modes, for each coefficient and history sum, so a run of $N$ steps costs $O(N^2)$; in the legacy solver, which keeps the load history on disk as text, each past increment is in addition re-read from file and re-transformed (Sect.~\ref{sec:implementation}).
With Eq.~(\ref{eq:recursion}), step $j$ performs one multiply--add per mode and coefficient, independent of $j$, and a run costs $O(N)$ from the first step.
The information carried between steps shrinks from the whole load history, 17\,GB for the 550 annual steps of Case 1 (Sect.~\ref{sec:cases}) to the $K\le26$ modal fields per degree and $\Delta L$ of 53\,MB.
This fixed-size state is also what a restart or a coupled ice sheet---GIA model has to establish a checkpoint.

\subsection{Placement in the nonlinear sea-level iteration}
\label{sec:placement}

\begin{figure*}[!tp]
\includegraphics[width=\textwidth]{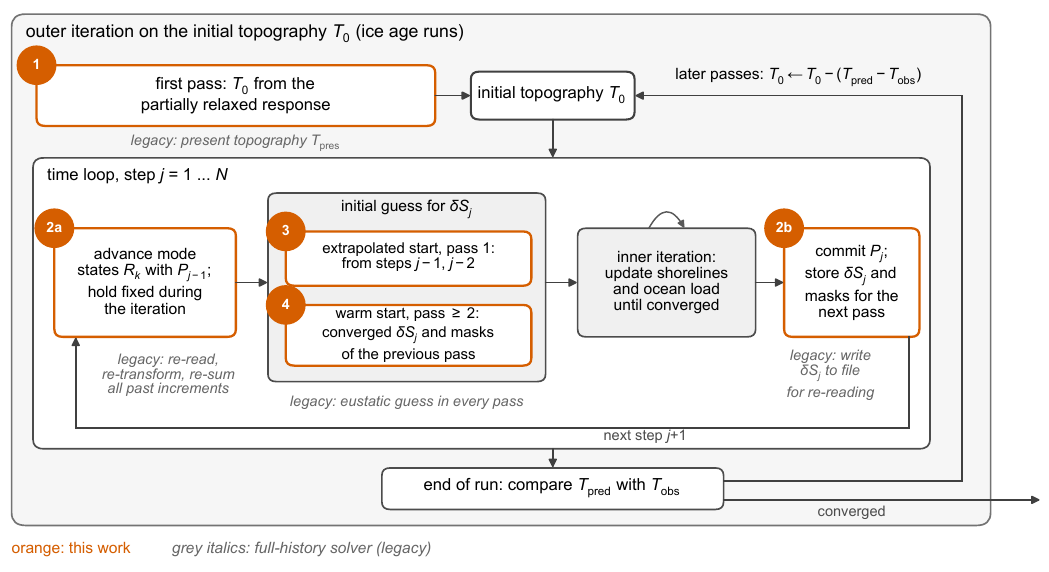}
\caption{Comparison between the proposed workflow in SLEvolve and that in the legacy solver.
The optimizations introduced here are numbered and highlighted in orange boxes: first-pass $T_0$ (1, Sect.~\ref{sec:paleo}), the state-update placement rule (2a, 2b, Sect.~\ref{sec:placement}), the extrapolated inner start (3, Sect.~\ref{sec:extrap}) and the warm start (4, Sect.~\ref{sec:paleo}). Gray italic texts describe what the legacy solver does instead.}
\label{fig:algorithm}
\end{figure*}

Because an increment applied at $t_j$ has zero viscous age at that time ($\beta_\ell(0)=0$), the viscous response at $t_j$ depends only on the converged cumulative load $\Delta L_{j-1}$ from the previous step. We exploit this property through a state-update placement rule: the modal states are advanced exactly once prior to entering the nonlinear inner Picard loop of step $j$, and remain fixed while the iteration updates the ocean function, marine grounding ice, and elastic sea-level increment. Trial iterates do not enter the stored history. After convergence, the load increment is committed:
$\Delta L_{\ell m}(t_j)=\Delta L_{\ell m}(t_{j-1})+\delta L^{j}_{\ell m}$
(Fig.~\ref{fig:algorithm}, steps 2a and 2b).

Because the response of a normal-mode Earth to a step increment is known in closed form over $\delta t$, no sub-iteration of the viscous evolution within the step is required. This differs from time-domain codes that use local stress as the memory state, which must estimate the strain rate within the step (Sect.~\ref{sec:discussion}). The inner iteration thus updates only the elastic and uniform terms, the ocean function and the grounding of marine ice, with the viscous contribution held fixed and the load represented as a step increment in time as in \citet{kendall2005}.

Advancing or committing the state inside the iteration would instead add unconverged trial loads to the history and alter the converged solution. With the placement rule above, each iteration sees exactly the viscous response of the full-history sum. Accordingly, the two solvers take identical numbers of inner iterations in all our tests (Sect.~\ref{sec:exactness}).

\subsection{Extrapolated inner start}
\label{sec:extrap}
The inner iteration of step $j$ starts from the linear extrapolation in time of the two previous converged increments (the superscript 0 marks the initial guess), $\delta S^{(0)}_j=\delta t_j\,[\,\dot S_{j-1}+(\dot S_{j-1}-\dot S_{j-2})\,(\delta t_j+\delta t_{j-1})/(\delta t_{j-1}+\delta t_{j-2})\,]$ with the rates $\dot S_i=\delta S_i/\delta t_i$. This reduces to $2\,\delta S_{j-1}-\delta S_{j-2}$ for uniform steps (the first two steps use the eustatic guess), instead of the eustatic guess of the legacy solver, which spreads the mass of the ice increment uniformly over the ocean. Because the increments vary smoothly from step to step, the extrapolation starts closer to the converged value.

\subsection{Initial topography of ice age runs}
\label{sec:paleo}

An ice age run requires the initial topography $T_0$, which is unknown.
It is found by an outer iteration that seeks $T_0$ such that the forward run ends at the observed present topography $T_{\mathrm{obs}}$.
We call each outer iteration a pass; an ice age calculation is a sequence of passes.
We use the Picard update of \citet{kendall2005},
\begin{equation}
T_0\leftarrow T_0-\left(T_{\mathrm{pred}}-T_{\mathrm{obs}}\right),
\label{eq:outer}
\end{equation}
where $T_{\mathrm{pred}}$ is the predicted present topography. This update converges because a change in $T_0$ produces nearly the same change in $T_{\mathrm{pred}}$, so subtracting the present-day topography error largely removes that error in $T_0$ in the next pass.
The misfit is the area-weighted global RMS (root mean square) of $T_{\mathrm{pred}}-T_{\mathrm{obs}}$; all RMS values in this paper are area-weighted over the globe unless a region is named.
We call the misfit after the three passes from $T_{\mathrm{obs}}$ used by \citet{kendall2005} the legacy accuracy.
The initial guess changes how many passes are needed, and usually the  present day topography is used as the initial guess of $T_0$ \citep{kendall2005}.
We consider two measures to reduce the cost of the outer iteration without changing its fixed point.

\subsubsection{Initial topography from the partially relaxed response}

There are two simple end members for a first-pass estimate of $T_0$. Using $T_0=T_{\mathrm{obs}}$ applies no GIA correction to the initial topography and is therefore the zero-correction baseline, whereas a fluid-limit correction assumes that the net load change has fully relaxed. Physics-based intermediate starts have also been used in 3-D GIA calculations. \citet{yuan2025} retain $T_0=T_{\mathrm{obs}}$ but prescribe the ocean-function history from a rigid-Earth barystatic sea level for runs that start at 122\,ka with ice close to present. For runs starting at 40\,ka, \citet{yuan2026} correct $T_0$ by local (fully compensated) isostasy of the ice-thickness change plus a barystatic level, the uniform sea-level change associated with the ice-volume change on a rigid Earth. We compute the latter over the present bathymetry and refer to this construction as the local-isostasy start.

Our partially relaxed start instead makes a simple, segment-based estimate of how much of the modeled loading history has actually relaxed. The history of global ice volume is split at its extrema into monotonic segments $i$ (changes smaller than 1\,\% of the volume range are ignored). For each segment, $\delta I_i$ is the net load change, defined as the grounded-ice field at the end of the segment minus that at its start, and $\tau_i$ is its mean age at the end of the run, weighted by the magnitude of the ice-volume change at each step. The response of each segment is then evaluated at this representative age using the actual degree-dependent viscoelastic Love-number response:
\begin{equation}
T_0=T_{\mathrm{obs}}+\sum_i\Delta\mathcal{SL}_i,\qquad h^{\tau_i}_\ell=h^E_\ell+\sum_k\frac{r_k^\ell}{s_k^\ell}\left(1-e^{-s_k^\ell\tau_i}\right),
\label{eq:tau}
\end{equation}
where $\Delta\mathcal{SL}_i$ is the sea-level response to $\delta I_i$ evaluated with the partially relaxed Love numbers $h^{\tau_i}_\ell$ and $k^{\tau_i}_\ell$ (the latter with the residues $r_k^{\ell\prime}$). The sea-level equation (Eq.~\ref{eq:sle}) is solved self-consistently once for the summed response. Thus, the approximation is only in replacing the detailed loading within each monotonic segment by its net change at a representative age, and in evaluating the water load with the segment-averaged Love numbers $h^{w}_\ell=\sum_i w_i\,h^{\tau_i}_\ell$ and $k^{w}_\ell=\sum_i w_i\,k^{\tau_i}_\ell$, $w_i=|\Delta V_i|/\sum_j|\Delta V_j|$, where $\Delta V_i$ is the net ice-volume change of segment $i$; the spatial and spectral response itself is that of the full 1-D Earth model. The construction requires no time stepping and has negligible cost relative to a forward pass.

Several segments are retained when the ice history contains distinct episodes of growth and decay, because load changes of different ages can nearly cancel in magnitude while leaving different viscoelastic responses. Only load changes within the modeled interval are included, consistent with the forward calculation in which the modal states start from zero at $t_0$. If the history is represented by a single segment, the fluid-limit start is recovered by taking $\tau\to\infty$; if the ice at $t_0$ equals the present ice and the history consists of a single segment, the partially relaxed guess reduces to $T_{\mathrm{obs}}$.

In Sect.~\ref{sec:iceage} we compare four first-pass choices: $T_{\mathrm{obs}}$, the fluid-limit start, the local-isostasy start, and the partially relaxed start above.

\subsubsection{Warm start across outer iterations}

The partially relaxed estimate improves the first outer pass. Subsequent passes offer a second opportunity for acceleration because they differ only through the updated initial topography $T_0$. At a given time step, the converged sea-level and ocean-mask fields from one pass are therefore generally close to those of the next pass and provide good initial guesses for its inner iteration.

We use this as a warm start for the inner iteration: the converged $\delta S_j$ and the ocean and grounded-ice masks ($C_j$ and $\beta_j$) at every step are stored and used to initialize the inner iteration at step $j$ in the next outer pass; the extrapolated start is used in the first pass and wherever no stored warm-start fields are available.
The storage involves one spectral field and two masks per step, about 5\,MB per step on the $512\times1024$ grid of the test cases.
The runs converge to the same fixed point within the prescribed tolerance (Sect.~\ref{sec:exactness}).

\subsection{Implementation}
\label{sec:implementation}

SLEvolve v1.0 contains a Fortran solver derived from the legacy code of \citet{han2022} and an independent Python version.
The recursive state is a self-contained module with two calls per step, advance-and-read before the inner iteration and commit after it, so that it can be inserted into other normal-mode solvers.
All solvers, including the legacy solver, run in double precision (fp64).
For each optimization, we state the bottleneck it removes and whether it alters the arithmetic or the convergence path.
We find techniques similar to several of the implementation and iteration-start methods developed and tested independently here are also employed in other sea-level codes, for example, in-memory execution with binary input and output, a precomputed Legendre table and warm starts in SL\_SLG\_TPW (\citealp{goldberg2016}; S. Goldberg, personal communication, 2026), additionally with the viscous kernel $\beta_\ell(t_j-t_n)$ tabulated for all pairs of time steps in SLcode \citep{austermann2015}, and batched parallel spherical harmonic transforms over all time steps in SELEN4 \citep{spada2019}.

We distinguish two complementary classes of optimization: (1) formulation-level complexity reduction, wherein the exact state-space recurrence reduces the asymptotic temporal complexity from $O(N^2)$ to $O(N)$; and (2) hardware-aware execution optimizations, which reduce constant-factor execution latency through resident in-memory states and binary I/O, thread-parallel Legendre transforms, and transform pruning. The two classes are not independent: the share of the history summation, and hence the gain of the recurrence, grows as the rest of the step becomes faster and as $N$ increases. While individual implementation measures (such as in-memory data structures or warm starts) have precedents in other codes, Table~\ref{tab:ablation} isolates each intervention to establish a staged ablation of cumulative speed-ups against the reference implementation of \citet{han2022}.

\subsubsection{Resident state and binary input and output (I/O)}
In the legacy solver, each time step is executed as a separate program run that re-reads and parses the ice files and ocean-load files of all previous steps as text. The full-history evaluation dominates the overall runtime, with text parsing its largest single cost (Sect.~\ref{sec:results}).
By contrast, SLEvolve advances all time steps in a single continuous run, keeping the modal states, Love numbers and step-0 fields in memory between steps. It also passes fields directly in the array layout expected by the spectral transform, avoiding copies between the legacy and transform layouts.
All grids are read and written in a binary floating-point format; the text files of the legacy code are converted once beforehand.

\subsubsection{Accelerated spherical-harmonic transforms}
On a Gauss--Legendre grid, a spherical-harmonic transform separates naturally into a Legendre transform in latitude and a Fourier transform in longitude.
The synthesis of a field from its coefficients $f_{\ell m}$ at colatitude $\theta_i$ and longitude $\phi_j$ is
\begin{equation}
f(\theta_i,\phi_j)=\sum_{m=-L}^{L}\Big[\sum_{\ell=|m|}^{L} f_{\ell m}\,P_{\ell m}(\cos\theta_i)\Big]e^{im\phi_j},
\label{eq:sht}
\end{equation}
where $L$ is the maximum spherical-harmonic degree (511 here). For each order $m$, the inner sum gives the latitudinal dependence through the associated Legendre functions $P_{\ell m}$, while the outer sum is a Fourier series in longitude.
Since the longitudes are equally spaced, $\phi_j=2\pi j/N_\phi$, the outer sum becomes a discrete Fourier transform of length $N_\phi$:
\begin{equation}
f(\theta_i,\phi_j)=\sum_{m}g_m(\theta_i)\,e^{2\pi i\,mj/N_\phi},\qquad g_m(\theta_i)=\sum_{\ell=|m|}^{L} f_{\ell m}\,P_{\ell m}(\cos\theta_i),
\label{eq:dft}
\end{equation}
An FFT (fast Fourier transform) evaluates this sum in $O(N_\phi\log N_\phi)$ rather than $O(N_\phi^2)$ operations. For $N_\phi=2^p$ ($N_\phi=1024$ here), the radix-2 algorithm splits the sum over even and odd $m$ into two transforms of length $N_\phi/2$, combines them using twiddle factors $e^{2\pi i\,j/N_\phi}$, and repeats the splitting recursively $p$ times \citep{cooley1965}. For real fields, Hermitian symmetry further reduces the Fourier work.
Analysis proceeds in the reverse order: an FFT along each latitude row gives the Fourier coefficients $g_m(\theta_i)$, followed by Gauss--Legendre quadrature over $\theta_i$ against $P_{\ell m}$ to recover $f_{\ell m}$.
In the legacy implementation \citep{han2022}, longitude transforms use FFT99 \citep{temperton1983}, while Legendre transforms use a precomputed table of $P_{\ell m}$ at all latitudes.

SLEvolve accelerates both parts. For the longitude transform, we replace FFT99, a scalar Fortran implementation with a fixed algorithm called separately for each latitude row, with FFTW \citep{frigo2005}. At start-up, the FFTW planner measures the runtime of candidate transform decompositions and selects a plan suited to the processor and memory hierarchy. The plan is created once for the full $512\times1024$ array and reused at every transform, allowing all latitude rows to be processed in a single call.

The larger gain, however, comes from the Legendre transform, which accounts for most of the transform cost. We implemented two backends. The first retains the precomputed $P_{\ell m}$ table but evaluates the Legendre sums as matrix multiplications using the Basic Linear Algebra Subprograms (BLAS; Apple Accelerate on the machine of Sect.~\ref{sec:verification}). It also exploits the north--south symmetry
\begin{equation}
P_{\ell m}(-\cos\theta)=(-1)^{\ell+m}P_{\ell m}(\cos\theta)
\label{eq:nssym}
\end{equation}
to halve the latitude work. This substantially reduces the cost on one or two threads, but repeatedly reading the large table makes the calculation memory-bandwidth limited and restricts its scaling with thread count.

The second backend avoids this bandwidth bottleneck by evaluating the associated Legendre functions on the fly using the standard recurrence in $m$ \citep{schaeffer2013}, without storing the table. The recurrence is vectorized over latitude and parallelized over $m$ with OpenMP, so its cost is dominated by arithmetic rather than memory traffic and it scales better with thread count. As a result, the table backend is faster at low thread counts, whereas the on-the-fly recurrence becomes faster from four threads onward (Sect.~\ref{sec:results}). SLEvolve therefore uses the on-the-fly backend by default.
Both Legendre backends and FFTW preserve the mathematical transform, changing only the order of floating-point operations. The final topographies of Case 1 differ by less than $10^{-11}$\,m between the backends (Sect.~\ref{sec:exactness}), and the output of each backend does not depend on the thread count.

\subsubsection{Fewer transforms per step}
Each inner iteration of the legacy step performs five spherical-harmonic transforms: analyses of the ocean load, the ocean function and the topography correction (the $T_0$ term of Eq.~\ref{eq:sle}), and syntheses of the sea-level increment and total sea-level change. Outside the loop, the step analyzes the ice and the first guesses of the ocean function and topography correction (three transforms), synthesizes the radial displacement (one), and analyzes four degree-0 diagnostics (global means for mass conservation and output); the two guesses are not re-analyzed after the last iteration, for a total of $5n_{\mathrm{iter}}+6$ transforms per step for $n_{\mathrm{iter}}$ inner iterations.
Two transforms per iteration are redundant: the total sea-level change is the increment plus a uniform term, and the two topography terms enter only as a difference, so one analysis serves both. Five additional transforms per step are unnecessary because the degree-0 diagnostics are grid means obtainable by quadrature and the radial displacement is needed only when it is written.
This leaves $3n_{\mathrm{iter}}+2$ transforms per step, with the combined guess now being analyzed after every iteration including the last.
The remaining analyses are grouped so that several fields share one evaluation of the Legendre sum, a Legendre sweep (one call for the ice and the first guess, one per iteration), so that the $3n_{\mathrm{iter}}+2$ transforms require only $2n_{\mathrm{iter}}+1$ Legendre sweeps; the arithmetic of each transform is unchanged.

\subsubsection{Python version}
We also created a Python version that keeps the modal state as a dense array of shape $K_{\max}\times(L+1)\times(L+1)$, with $K_{\max}=\max_\ell K_\ell$, and delegates the transforms to ducc0 \citep{reinecke_ducc0}, a threaded spherical-harmonic transform library for Gauss--Legendre grids that needs no precomputed table, and compiles the recurrence and the spectral assembly with Numba \citep{lam2015}.

\subsection{AI-assisted implementation}
\label{sec:ai}

SLEvolve was developed through a human-directed, AI-assisted engineering workflow guided by a benchmark harness. The authors derived the recursive-state formulation, proposed optimization directions and defined numerical acceptance criteria. AI coding agents (Claude Fable 5.1 and Claude Opus 5.5) implemented, optimized, ported and tested the code. The authors verified the results and selected the methods retained in the final solver.

The harness applied two tiers of numerical verification. Tier 1 (machine-precision agreement): changes that did not alter iteration starts, including the recursive formulation, memory reorganization and transform refactoring, were required to reproduce the legacy solver to machine precision (pointwise topography differences below $10^{-11}$\,m), with identical inner-iteration counts at every step. Tier 2 (converged-solution agreement): changes that altered iteration starts, including extrapolation, warm starts and the partially relaxed initial topography, were required to converge to the same solution within the prescribed tolerance ($\epsilon=10^{-5}$; Sect.~\ref{sec:exactness}).

Candidate optimizations were also judged by total runtime. Anderson acceleration, secant outer updates and inter-pass warm-start adjustments were tested but not retained because their overhead outweighed their benefits or they did not reduce the number of passes at the target tolerances (Sect.~\ref{sec:discussion}).

\section{Cases and timing}
\label{sec:verification}

\subsection{Cases}
\label{sec:cases}
The three test cases comprise one forward (projection) run and two ice age runs, all on the $512\times1024$ Gauss--Legendre grid of \citet{han2022} (Table~\ref{tab:cases}).
Case 1 is the Antarctic case of \citet{han2022}: 550 annual steps from 1950 CE, driven by the total (grounded plus floating) ice of the \citet{deconto2021} Antarctic ice-sheet simulation used in that study, recovered from the archive of the same simulation by \citet{yousefi2022,yousefi2022data}. This forcing reproduces the archived result of \citet{han2022}.
Case 2 is the two-glacial-cycle benchmark of \citet[][their Sect.~3.3.1]{han2022}: 1200 steps of 0.2\,kyr from 240\,ka with the global ice history distributed with that study \citep{han2021osf}. The history combines a Northern Hemisphere ice-sheet simulation \citep{han2021}, repeated over two cycles, with ICE-6G\_C for Antarctica \citep{argus2014,peltier2015}. For Case 2, we report only the first outer pass, initialized from $T_{\mathrm{obs}}$, for the timing comparison between the legacy solver and SLEvolve for the full 240 kyr of two glacial cycles. For the comparison of different initial-$T_0$ guesses, however, we run the outer iteration to convergence for each starting time within the last glacial cycle (Sect.~\ref{sec:iceage}).
Case 3 is the ICE-NADI-1 deglaciation \citep{lucas2026,lucas2026data}, 50 steps of 0.5\,kyr from 25\,ka, with the outer iteration on $T_0$.

\begin{table*}[!tp]
\caption{The three test cases considered in this study, all on the $512\times1024$ Gauss--Legendre grid. PREM: Preliminary Reference Earth Model \citep{dziewonski1981}; the lithosphere is represented as a layer of effectively infinite viscosity.}
\label{tab:cases}
\footnotesize\setlength{\tabcolsep}{4pt}\renewcommand{\arraystretch}{1.1}
\begin{tabular}{@{}l>{\raggedright\arraybackslash}p{1.6cm}>{\raggedright\arraybackslash}p{4.4cm}>{\raggedright\arraybackslash}p{2.6cm}>{\raggedright\arraybackslash}p{4.4cm}>{\raggedright\arraybackslash}p{2.2cm}@{}}
\tophline
Case & Run type & Ice history model & Steps & Earth model & Outer iteration on $T_0$\\
\middlehline
1 & forward & an extended RCP8.5 (Representative Concentration Pathway) MISI-only (Marine Ice Sheet Instability) run of \citet{deconto2021}, ice file released in \citet{yousefi2022} & 550 $\times$ 1\,yr from 1950\,CE & PREM (elastic and density profiles); viscosity: 60-km lithosphere, upper mantle $10^{18.6}$--$10^{19.4}$\,Pa\,s (depth-dependent), lower mantle $5\times10^{22}$\,Pa\,s in \citet{han2022}, modified from \citet{barletta2018,deconto2021} & not needed (forward run)\\
2 & ice age & global ice sheets over two glacial cycles \citep{han2022}: Northern Hemisphere simulation of \citet{han2021} with ICE-6G\_C Antarctica \citep{peltier2015} & 1200 $\times$ 0.2\,kyr from 240\,ka & PREM (elastic and density profiles); viscosity: 120-km lithosphere, upper mantle $5\times10^{20}$, lower mantle $5\times10^{21}$\,Pa\,s \citep{han2022} & first pass only\\
3 & ice age & ICE-NADI-1 \citep{lucas2026} & 50 $\times$ 0.5\,kyr from 25\,ka & PREM (elastic and density profiles); viscosity: 96-km lithosphere, upper mantle $3\times10^{20}$, lower mantle $10^{21}$\,Pa\,s \citep{lucas2026} & complete (up to 8 passes)\\
\bottomhline
\end{tabular}
\end{table*}

\subsection{Reference solver and timing protocol}
We use the legacy solver of \citet{han2022} as our primary reference because it is the most recent publicly released implementation in this 1-D GIA code lineage summarized in Table \ref{tab:lineage} and supports both forward projections and ice-age calculations \citep{gomez2010,han2022}. All timings were measured on an Apple M4 Max with 12 performance cores, 4 efficiency cores and 128\,GB unified memory.
Each SLEvolve configuration was run three times, and the median runtime is reported; the legacy solver was run once per benchmark because of its high computational cost.
We also report the performance of other full-history implementations, like SLcode \citep{austermann2015}, SL\_SLG\_TPW (\citealp{goldberg2016}; S. Goldberg, personal communication, 2026), SELEN4 \citep{spada2019} and giapy \citep{kachuck2019}, some of them are faster than \citet{han2022} and some are slower.
\section{Results}
\label{sec:results}

\subsection{Exactness}
\label{sec:exactness}
\begin{figure*}[!tp]
\includegraphics[width=\textwidth]{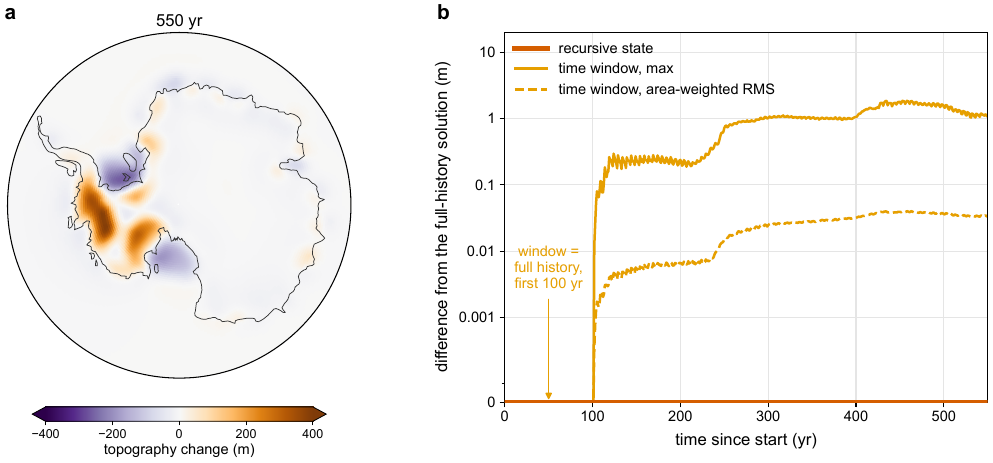}
\caption{(a) Topography change since $t_0$ at 550\,yr in Case 1, recursive-state run.
(b) Difference in topography from the legacy solution against time for the same case: the time window of \citet{han2022} (maximum and area-weighted RMS) is exactly zero for the first 100\,yr while all increments remain at annual resolution, whereas the difference of the recursive-state solution from the legacy solution is below plot resolution throughout (drawn as 0).}
\label{fig:verification}
\end{figure*}

In all three cases, with the same eustatic inner start and initial topography as the legacy solver, SLEvolve reproduces the legacy solution to machine precision: the fields differ by less than $10^{-11}$\,m, consistent with floating-point round-off, and the ocean and grounded-ice masks and numbers of inner iterations are identical at every step (Table~\ref{tab:ablation}).
This agreement holds across the implementation measures of Sect.~\ref{sec:implementation}, tested thread counts and the independent Python version.
The extrapolated and warm inner starts and the partially relaxed $T_0$ alter the initial guesses, but the calculations converge to the same solution within the prescribed tolerance.
By contrast, the time window of \citet{han2022}, which retains the most recent 100 increments at annual resolution and merges older increments, is exact relative to the full-history solver only for the first 100 steps. Once earlier increments are coarsened, the result differs from the legacy solution by only $\sim$1\,m at 550\,yr, or 0.2\,\% of the deformation (Fig.~\ref{fig:verification}), consistent with \citet{han2022}.

\subsection{Projection run: cost and scaling}
To deconstruct the compound acceleration, Table~\ref{tab:ablation} presents a staged ablation isolating the contributions of formulation complexity, hardware-aware refactoring and iterative workflow acceleration. Algorithmic complexity reduction (Stage 1): replacing the legacy convolution sum with the modal recurrence reduces the asymptotic scaling from $O(N^2)$ to $O(N)$ and shortens the Case 1 run from 24\,309\,s to 1\,180\,s (20.6-fold) while operating within the legacy file-based harness; most of this gain comes from no longer replaying the history from files, since an in-memory full-history solver of the same scheme already runs in 870\,s (Sect.~\ref{sec:iceage}), and the recurrence itself contributes a factor of about 1.5 at $N=550$ that grows with $N$. I/O and memory hierarchy modernization (Stage 2): eliminating disk round trips through resident in-memory states, binary I/O and the transform interface provides an additional 4.7-fold gain (97-fold cumulative). Spectral transform and arithmetic pruning (Stages 3--5): the Legendre table as a matrix product with north--south symmetry, FFTW with shared Legendre sweeps, and the elimination of redundant transforms compound a further 4.1-fold improvement, driving the serial speed-up to 402-fold. Fixed-point iteration starts (Stage 6): the extrapolated start reduces the inner iterations per step from 4.6 to 3.8, pushing the serial acceleration to 464-fold.

\begin{table*}[!t]
\caption{Stages of the implementation on Case 1, the 550-step future projection of \citet{han2022}. (a) One thread; each row cumulatively adds the measures listed in that row to the row above. Gain is the incremental speed-up relative to the preceding row, whereas Speed-up is the cumulative speed-up relative to the legacy solver. (b) Multi-threading results for the two Legendre backends and the Python version, all with the extrapolated inner start; parentheses give the speed-up over the legacy run and the parallel efficiency, defined for each backend as $T_1/(pT_p)$, where $T_p$ is the wall time on $p$ threads. Wall times of the recursive builds are medians of three runs on an otherwise idle machine; inner iterations are means per time step. Speed-ups are relative to the legacy implementation of \citet{han2022}; some of the optimizations excluding recursive state are also similarly implemented in other sea-level codes (Sect.~\ref{sec:implementation}).}
\label{tab:ablation}
\footnotesize\setlength{\tabcolsep}{2.5pt}
\begin{tabular}{rlrrrrr}
\tophline
\multicolumn{7}{l}{(a) One thread, precomputed Legendre table}\\
\middlehline
 & Stage & Wall (s) & Gain & Speed-up & Peak memory (GB) & Inner iter./step\\
\middlehline
0 & legacy (full-history solver, text file) & 24\,309 (6.8 h) & -- & 1 & 27.9 & 4.6\\
1 & recursive state, otherwise legacy code & 1\,180 & 20.6 & \textbf{20.6} & 3.01 & 4.6\\
2 & resident state, binary I/O, transform interface & 250.9 & 4.70 & 97 & 1.60 & 4.6\\
3 & Legendre table as matrix product, N--S symmetry & 163.0 & 1.54 & 149 & 1.05 & 4.6\\
4 & FFTW, shared Legendre sweeps & 103.4 & 1.58 & 235 & 1.02 & 4.6\\
5 & fewer transforms per step ($5n{+}6 \rightarrow 3n{+}2$) & 60.5 & 1.71 & 402 & 1.02 & 4.6\\
6 & extrapolated inner start & 52.4 & 1.16 & \textbf{464} & 1.03 & 3.8\\
\bottomhline
\end{tabular}

\medskip
\begin{tabular}{rlrrrrrr}
\tophline
\multicolumn{8}{l}{(b) Multi-threading: wall time in s (speed-up over the legacy run; parallel efficiency)}\\
\middlehline
 & Backend & 1 thread & 2 threads & 4 threads & 8 threads & 12 threads & Peak memory (GB)\\
\middlehline
6 & precomputed Legendre table & 52.4 (464; 100\,\%) & 31.8 (764; 82\,\%) & 24.6 (990; 53\,\%) & 19.0 (1279; 34\,\%) & 16.6 (1467; 26\,\%) & 1.03\\
7 & on-the-fly Legendre transform & 58.7 (414; 100\,\%) & 33.5 (727; 88\,\%) & 20.2 (1204; 73\,\%) & 13.1 (1851; 56\,\%) & 10.9 (2222; 45\,\%) & 0.51\\
7p & Python version of row 7 & 70.1 (347; 100\,\%) & 39.6 (614; 89\,\%) & 23.6 (1031; 74\,\%) & 16.5 (1475; 53\,\%) & 14.0 (1731; 42\,\%) & 0.70\\
\bottomhline
\end{tabular}
\belowtable{Rows 1--5 reproduce the legacy solver to machine precision with the same number of inner iterations at every step (Sect.~\ref{sec:exactness}); rows 6--7p start the inner iteration from an extrapolated guess and converge within the same tolerance.}
\end{table*}

In Case 1, the legacy 550-step run takes 24\,309\,s (6.8\,h), compared with 12\,249\,s (3.4\,h) using the time-window approximation. Most of the legacy runtime (89\,\%) is spent replaying history from files: the run parses about 3.2\,TB of text. Parsing accounts for 67\,\% of the total runtime, redundant transforms for 21\,\%, and file reads for less than 1\,\%.

\begin{figure*}[!tp]
\includegraphics[width=\textwidth]{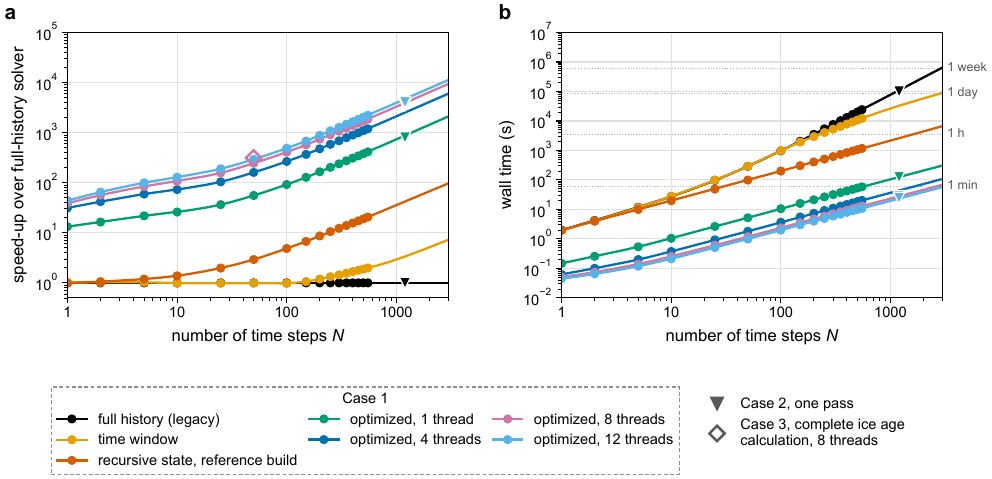}
\caption{(a) Speed-up over the legacy solver and (b) wall time against the number of time steps $N$ for Case 1 (Table~\ref{tab:ablation}).
Dots are the measured cumulative wall times after the first $N$ steps of the 550-step runs; lines connect the measurements and, beyond $N=550$, are extrapolated using their terminal slope in log--log coordinates.
The time window of \citet{han2022} was run on the same case with the same inputs; beyond $N=550$ its line is continued with the measured cost of its last steps, assuming that the number of sampled past steps does not increase further.
Optimized SLEvolve runs use the on-the-fly Legendre transform.
Triangles mark one pass of Case 2 (Table~\ref{tab:glacial}), whose ice history and Earth model (smaller ice files, fewer relaxation modes) make each past step cheaper for the legacy solver.
The diamond marks the complete ice age calculation of Case 3 on 8 threads to the legacy accuracy (Sect.~\ref{sec:paleo}; Table~\ref{tab:paleo}).
Dotted lines in (b) mark 1 min, 1 h, 1 day and 1 week.}
\label{fig:scaling}
\end{figure*}

The recursive state alone reduces the run to 1180\,s (20.6 times faster). At year 550, a single step is 39 times faster; the cumulative speed-up grows approximately linearly with $N$ because the legacy run costs $O(N^2)$ whereas the recursive run costs $O(N)$ (Fig.~\ref{fig:scaling}). For reference, the inner iteration takes 2 to 6 iterations per step in Case 1, averaging 4.6 (Table~\ref{tab:ablation}).

The implementation measures of Sect.~\ref{sec:implementation} are independent of $N$ and multiply this gain (Table~\ref{tab:ablation}). On 8 threads, the final Fortran version takes 13.1\,s (1851 times faster) and uses 0.5\,GB of peak memory, compared with 27.9\,GB for the legacy solver. The table also reports the 1- and 4-thread timings, the precomputed-table backend, and the Python version, whose runtime is within 30\,\% of the Fortran version.

The two Legendre backends have different threading behavior. The precomputed table is faster on one and two threads (52.4 versus 58.7\,s and 31.8 versus 33.5\,s), whereas the on-the-fly transform is faster from four threads onward and uses about half the memory. At 12 threads, equal to the number of performance cores of the M4 Max used here, the parallel speed-up begins to saturate: runtimes fall to 10.9\,s in Case 1 (2222 fold speed-up instead of 1851), approximately 17\,\% shorter than on 8 threads.

The extrapolated inner start alone requires 17\,\% fewer inner iterations in Case 1. (Analogous numbers are 21\,\% in Case 2 and 4\,\% in Case 3, not tabulated). For 5000 time steps, although not directly tested, extrapolating the measured quadratic legacy scaling gives about $1.8\times10^{6}$\,s (three weeks), whereas extrapolating the linear scaling of the final 8-thread solver from the measured 550-step run gives about 120\,s.

\subsection{Ice age runs: one pass and outer iteration}
\label{sec:iceage}

Ice age runs repeat the forward run in an outer iteration on the initial topography, so every pass of the legacy solver incurs the $O(N^2)$ history cost.
In Case 2, with 1200 steps of 0.2\,kyr, the legacy solver needs 100\,087\,s (27.8\,h) per pass. SLEvolve instead takes 27.9\,s on 8 threads and 24.7\,s on 12 threads, a speed-up of more than 4000 on 12 threads for a single outer iteration (Table~\ref{tab:glacial}).


\begin{table*}[!tp]
\caption{Cost of one pass of Case 2, the 1200-step glacial cycle of \citet{han2022}, on the $512\times1024$ grid. The legacy solver is run on one thread. The timings shown here are for the first pass of the outer iteration, i.e., a complete forward run through the 240-kyr history.}
\label{tab:glacial}
\footnotesize\setlength{\tabcolsep}{6pt}
\begin{tabular}{lrr}
\tophline
Code & Wall time (s) & Speed-up\\
\middlehline
legacy, 1 thread & 100\,087 (27.8\,h) & 1\\
SLEvolve, 1 thread & 124.8 & 802\\
SLEvolve, 8 threads & 27.9 & 3582\\
SLEvolve, 12 threads & 24.7 & \textbf{4054}\\
\bottomhline
\end{tabular}
\belowtable{A complete ice age calculation gains further from the partially relaxed initial-topography guess and the warm start (Sect.~\ref{sec:paleo}).}
\end{table*}

\begin{figure*}[!tp]
\includegraphics[width=\textwidth]{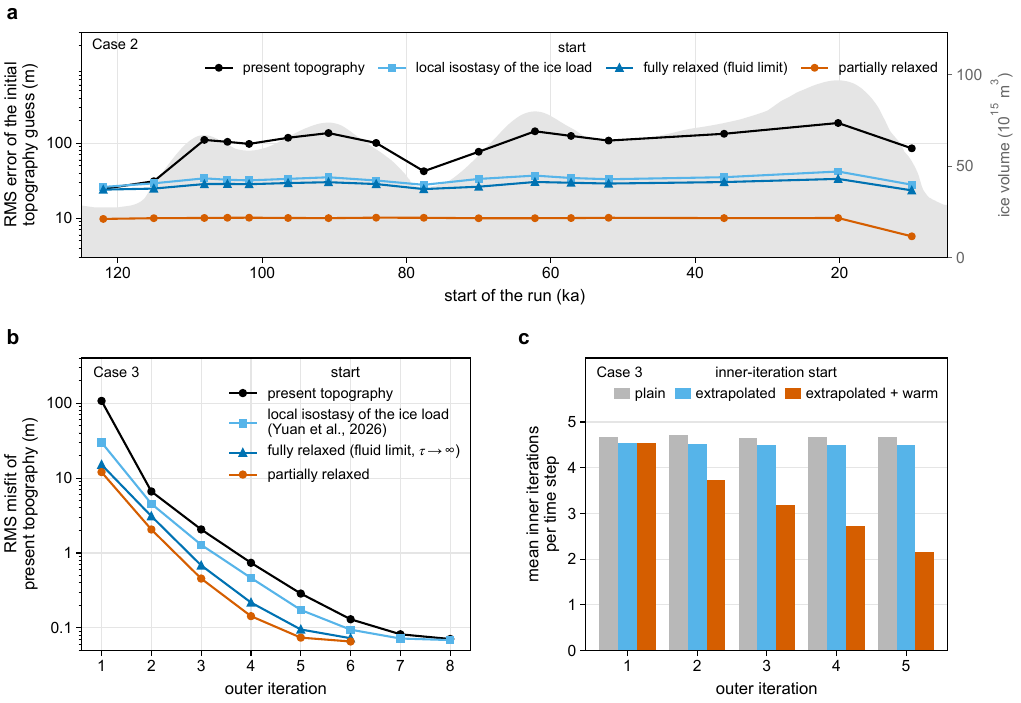}
\caption{Initial topography and outer iteration of ice age runs. (a) RMS error of the initial-topography guess relative to the converged initial topography, for runs starting at different times in the last glacial cycle of the Case 2 history (gray: ice volume); the partially relaxed guess stays near 10\,m at every start, an order of magnitude below the present topography at glacial starts. (b) RMS misfit of the predicted present topography against outer iteration in Case 3 for four starting topographies (Sect.~\ref{sec:paleo}), with the eustatic inner start of the legacy solver and no warm start, so that only the effect of the initial topography is shown. (c) Mean inner iterations per time step in each outer iteration of Case 3 from the partially relaxed start, for the eustatic, extrapolated and extrapolated plus warm inner starts; the misfit is effectively unchanged across the three runs, so only the work per pass changes.}
\label{fig:iceage}
\end{figure*}

Figure~\ref{fig:iceage}a compares the four initial-topography starts over the last glacial cycle of the Case 2 history, using runs starting every 10 to 20\,kyr between 122 and 10\,ka. Each guess is evaluated against the initial topography to which that run converges. At glacial starts, when ice volume is relatively large (110 to 10\,ka), starting from the present topography gives an RMS error of 62 to 186\,m, whereas the local-isostasy and fluid-limit starts leave 23 to 42\,m and the partially relaxed guess only 6 to 10\,m. At the interglacial start (122\,ka), where ice volume is close to the present value, the other starts leave 24 to 26\,m, compared with about 10\,m for the partially relaxed guess. This is about an order of magnitude below the present-topography start at the glacial starts and a factor of 2.5 lower at the interglacial start.

The partially relaxed guess is the best start at every start time tested. Its advantage comes from retaining finite, degree-dependent relaxation rather than either end member. A load change of age $\tau$ is almost fully relaxed at the ice-sheet scales (degrees above about 16) once $\tau$ exceeds 10 to 20\,kyr, whereas the lowest degrees are still 10 to 20\,\% short of the fluid limit after 80\,kyr. The present-topography start therefore neglects relaxation that occurs within the run, while the fluid-limit start over-relaxes the longest wavelengths. The partially relaxed construction instead evaluates each segment at its own representative age using the same Love-number spectrum as the forward model. Relative to the fluid-limit start, its improvement is a factor of about 3 in Case 2 but only 1.3 in Case 3 (Fig.~\ref{fig:iceage}a,b). This is consistent with the five-times-higher lower-mantle viscosity in the Case 2 model (Table~\ref{tab:cases}), for which the lowest degrees remain farther from isostatic equilibrium within the run. The remaining error is concentrated under the ice at $t_0$ (20 to 34\,m); the ice-free error is 3 to 9\,m.

The segment construction also matters when the ice history contains more than one growth or decay phase. In the ICE-NADI-1 history of Case 3 (Sect.~\ref{sec:cases}), a short growth segment from 25 to 21\,ka ($\tau=23$\,kyr) is followed by the main deglaciation from 21 to 0\,ka ($\tau=12.5$\,kyr). Earlier loading and unloading can nearly cancel in net ice change but not in response because the two contributions have different ages. As a consistency test on the assumed start of the modeled memory, prefixing Case 3 with 15\,kyr of constant glacial-maximum ice (40 to 25\,ka) leaves the converged $T_0$ unchanged, because no load increments are added during that interval.

Case 3 has only 50 steps, so the history term is a smaller share of each pass. For the different initial-topography guesses in Case 3, the misfit falls along roughly parallel curves (Fig.~\ref{fig:iceage}b): a better initial guess lowers the misfit of each pass, while all starts converge to the same $T_0$. The partially relaxed start lowers the first-pass misfit from 107\,m for $T_{\mathrm{obs}}$ to 12.0\,m, compared with 15.1\,m for the fluid-limit start and 29.9\,m for local isostasy. The local-isostasy guess retains sharp ice-margin features that are absent from the Earth's viscoelastic response, and its rigid-Earth barystatic term overshoots because the self-consistent pattern already includes about 25\,m of mean ocean-floor deformation. The partially relaxed start reaches the legacy accuracy (about 2\,m RMS) in two passes instead of three, and 0.1\,m RMS in five passes instead of seven (Table~\ref{tab:paleo}).

\citet{yuan2026} suggest that a single pass is sufficient based on relative sea-level histories at sites, which in the far field are controlled mainly by ice volume. The global present-topography misfit used here is instead most sensitive to $T_0$ beneath and near the former ice sheets, where the local-isostasy approximation contributes most of its error; the misfit therefore continues to decrease over further passes.

The warm start further reduces the cost of these passes, lowering the mean number of inner iterations per time step from 4.5 in the first pass to 3.7, 3.2, 2.7 and 2.2 in passes 2 to 5, a reduction of up to 52\,\% (Fig.~\ref{fig:iceage}c). The complete workflow takes 3.09\,s on 8 threads, 313 times faster than the legacy solver (Table~\ref{tab:paleo}).

\begin{table*}[!tp]
\caption{Complete ice age calculation of Case 3 with the outer iteration on the initial topography $T_0$. Legacy starts from $T_{\mathrm{obs}}$; SLEvolve starts from the partially relaxed guess (Eq.~\ref{eq:tau}) with extrapolated inner start and warm start. Accuracy is the RMS misfit of the predicted present topography (Sect.~\ref{sec:paleo}); legacy accuracy is the misfit that the full-history solver reaches after the three outer iterations from $T_{\mathrm{obs}}$ used by \citet{kendall2005}, about 2\,m here. Entries give the number of passes and the total time, with the speed-up over the legacy calculation in brackets; times of SLEvolve are medians of three runs.}
\label{tab:paleo}
\footnotesize\setlength{\tabcolsep}{4pt}
\begin{tabular}{lllll}
\tophline
Code & Time per pass (s), pass 1 $\rightarrow$ 5 & To legacy accuracy ($\approx$2\,m) & To 1\,m & To 0.1\,m\\
\middlehline
legacy & 321.9 & 3 passes: 966\,s & 4 passes: 1287\,s & 7 passes: 2253\,s\\
SLEvolve, Fortran & 1.73 $\rightarrow$ 1.02 & 2 passes: \textbf{3.09\,s (313$\times$)} & 3 passes: 4.33\,s (297$\times$) & 5 passes: 6.49\,s (347$\times$)\\
SLEvolve, Python & 1.68 $\rightarrow$ 1.01 & 2 passes: 3.12\,s (309$\times$) & 3 passes: 4.43\,s (291$\times$) & 5 passes: 6.60\,s (341$\times$)\\
\bottomhline
\end{tabular}
\belowtable{SLEvolve on 8 threads; on 12 threads SLEvolve can be about 15\,\% faster; legacy on one thread.}
\end{table*}

\subsection{Comparisons with other codes}
Note that when we compare with SL\_SLG\_TPW (\citealp{goldberg2016}; S. Goldberg, personal communication, 2026) that incorporates in-memory execution with binary input and output, a precomputed Legendre table, and warm starts, SLEvolve 12 threads still achieves 80 and 105 fold of speed-up for case 1 and 2, respectively (SL\_SLG\_TPW is 28 and 39 times faster than the \citep{han2022} for case 1 and 2, respectively). Even just replacing only the history summation of SL\_SLG\_TPW by the recursive state, with everything else unchanged, makes it about twice as fast in both cases. SLcode \citep{austermann2015}, which has the same in-memory full-history design, is about 4.8 and 3.0 times slower than SL\_SLG\_TPW for Case 1 and Case 2 respectively, mainly from the difference in the Fortran (SL\_SLG\_TPW) and Matlab (SLcode) backend. The original \citet{kendall2005} code is slightly slower than SL\_SLG\_TPW but they have overall comparable performance.

We also find giapy \citep{kachuck2019}, a Python solver without rotational feedback, is about 16 and 25 times faster than the legacy solver in \citet{han2022} for case 1 and 2, respectively. Another code SELEN4 \citep{spada2019} that supports parallel processing takes 1456 s and 3022 s on 12 threads (11 500 s and 25 006 s on one thread), but at a resolution equivalent to degree 128, a four times lower spectral truncation than ours. Scaling to the degree-512 setting would take about 33 h and 69 h on 12 threads (262 h and 569 h on one thread), about 5 and 2.5 times (39 and 20 times on one thread) slower than the legacy solver, and would require more memory than available on our laptop (M4 Max, 128 GB).

\section{Discussion}
\label{sec:discussion}

The quadratic cost of normal-mode GIA solvers is a property of how the loading history is represented, not of the physics. The full-history formulation re-sums past load increments at every step; the recursive formulation carries their combined response in a finite-dimensional state-space representation. Because the kernel is a finite sum of exponentials, the recurrence in Eq.~(\ref{eq:recursion}) is the exact update for the prescribed step-wise loading. The two formulations are algebraically identical, and the history-summation cost is reduced from $O(N^2)$ to $O(N)$ without coarsening the load history.

The recursive normal-mode state and the stress state of a time-domain code differ in their driving variables.
In a time-domain code the state is the stress field, and advancing it over $\delta t$ requires the strain rate within the step, which is part of the unknown solution.
Here the state is driven by the load, which is prescribed as a step-wise load in the discretization of \citet{kendall2005} used by both the full-history and recursive solvers.
For this prescribed load history, the normal modes give the exact viscoelastic response, so the step update $e^{-s_k^\ell\delta t}$ introduces no time-integration error.
The same state formulation also applies when the prescribed load varies within a step: the increment $\delta L^{\,j-1}$ in Eq.~(\ref{eq:qrec}) is then replaced by the integral of the load path over the step weighted by $e^{-s_k^\ell(t_j-t')}$.
For piecewise-linear or piecewise-polynomial loading, this integral can be evaluated analytically at essentially the same cost per step.

The recursive state therefore does not transfer directly to time-domain 3-D codes, which already carry the history in their mechanical state.
The three workflow measures, however---the extrapolated inner start, the warm start and the partially relaxed initial topography (Sects.~\ref{sec:extrap} and \ref{sec:paleo})---do not depend on the normal-mode representation and target the expensive iterative parts of such codes.
Within a step, CitcomSVE iterates between a multigrid solve for the incremental displacement and an update of the incremental potential until convergence, starting every step from zero potential forcing \citep{yuan2025}. An extrapolated start from the two previous increments could therefore reduce the number of 3-D solves.
SeaKon \citep{latychev2005} iterates at each time step between gravity and displacement with the generalized minimal residual (GMRES) linear solver \citep{saad1986}. During these iterations, the displacement iterates are used to advance the viscoelastic memory, with several iterations per viscoelastic state, following a modified form of the exponential propagator of \citet[][see also the supporting information of \citealp{lau2026}]{bailey2006}. For a single mode, this propagator has the same structure as Eq.~(\ref{eq:recursion}). The extrapolated increments and stored masks could therefore be used to initialize these iterations.
VILMA, by contrast, advances the sea-level equation explicitly with the time step using the ocean load of the previous step \citep{hagedoorn2007,martinec2018} and therefore has no within-step sea-level fixed-point iteration to which the extrapolated start could be applied.

Ice age calculations with these codes also require repeated passes through the complete loading history to determine the initial topography, with up to six passes required to reach a topography tolerance of 0.01\,m in the benchmark of \citet{martinec2018}. The computational cost of each complete run can be substantial: \citet{yuan2025} report 3\,h on 384 cores for a single 122-kyr glacial-cycle simulation with CitcomSVE-3.0 at approximately 50\,km horizontal resolution (about 13 million nodes), while \citet{love2024} report approximately 1.75 core-years (or equivalently 40 h on 384 cores) for a typical glacial-cycle SeaKon simulation, with approximately 15\,km at the surface (about 17 million nodes). With more recent computing clusters, SeaKon is now about twice as fast, and takes about 7 min per step for 25 million nodes on 384 cores. Although these benchmarks use different numerical settings and are not directly comparable, they illustrate the substantial computational savings that could result from reducing the number of outer passes in 3-D GIA calculations.
Storing the converged increments and masks at every step for the next pass reduces the number of inner iterations in our later passes (Sect.~\ref{sec:iceage}), and this warm start strategy could be used to initialize the iterative solvers at each step in all these 3-D codes.
The partially relaxed initial-topography guess could similarly reduce the number of complete passes by providing a physics-based first estimate before the expensive 3-D outer iteration; a converged 1-D solution of the kind used to initialize 3-D calculations by \citet{gomez2018} now takes seconds to minutes.

The construction applies to any solver whose viscous kernel is a finite sum of exponentials.
This includes normal-mode and collocation Love numbers for spherically symmetric Maxwell Earth models, as well as linear rheologies with additional discrete relaxation times, such as the Burgers body of \citet{caron2017} or a generalized Maxwell body.
Transient rheologies with a continuous distribution of relaxation times, such as the extended Burgers and related anelastic laws \citep{lau2021,ivins2022}, have no finite modal expansion.
In practice, however, their Love numbers can be represented by a finite set of fitted collocation poles \citep{mitrovica1992,lau2023,lau2024}; the resulting exponential sums have been used in the same 1-D pseudo-spectral solver considered here \citep{lau2023,coonin2025} and, in the form of a ten-branch generalized Maxwell body, in a 3-D code \citep{lau2026}.
In this case, the recurrence is exact for the fitted kernel, and the fitting procedure is the only approximation.
Love numbers computed in the time domain, as in giapy \citep{kachuck2019}, can likewise be represented by an exponential fit before applying the recurrence.
Nonlinear, stress-dependent creep is outside the construction, because the response is then not a superposition of load increments.
Laterally varying viscosity couples spherical-harmonic degrees, so the response is no longer a finite exponential sum independently at each degree; a degree-by-degree recurrence of the present form would therefore be approximate.

The history-independent per-step cost makes full gravitationally self-consistent 1-D GIA more practical for coupled ice-sheet simulations, large ensembles, uncertainty quantification, and repeated forward evaluations in inversions. Adjoint sensitivity methods have already been developed for gravitationally self-consistent GIA, including ice-history and mantle-viscosity sensitivities, using spatially distributed viscous stress \citep{martinec2015} or strain-like Maxwell internal variables \citep{crawford2018,yu2025} to represent viscoelastic memory. The fixed-size modal state provides an alternative basis for a future discrete adjoint of the 1-D pseudo-spectral sea-level solver, in which sensitivities could be propagated backward through the transpose of the modal state-update operator. Implementing this extension in SLEvolve would require treatment of the nonlinear sea-level iteration and changing shorelines, storage or reconstruction of forward states, and, for mantle-viscosity sensitivities, derivatives of the Love-number relaxation rates and residues with respect to Earth parameters. Such an adjoint could enable gradient-based inversion constrained by the GIA forward model, particularly for high-dimensional ice-history parameterizations.

Fast, physics-based forward calculations also make it practical to generate training sets spanning ice histories and radial Earth models. GIA emulators have already demonstrated the value of this approach, including GEORGIA, a graph-neural-network emulator trained on outputs from a gravitationally self-consistent 1-D GIA model \citep{lin2023}, and a neural-network surrogate for 3-D GIA calculations \citep{love2024}. Such training sets could also support operator-learning approaches, including Fourier neural operators (FNOs) \citep{li2021fno} and deep operator networks (DeepONets) \citep{lu2021deeponet}, to learn mappings from ice-loading histories and Earth parameters to sea-level and deformation fields. These are potential downstream applications, not capabilities implemented or benchmarked in SLEvolve.

The fixed-size state simplifies restarts and checkpointing in coupled systems. When a coupled or adaptive time step is rejected, however, the state must be committed only after acceptance or saved so that it can be restored.

The human-derived recurrence provided the main reduction in computational complexity, while AI coding agents explored implementation and workflow alternatives to reduce the remaining cost. A benchmark harness combining numerical regression against the legacy solver with runtime profiling guided the selection of optimizations. Anderson acceleration \citep{walker2011} saved some inner iterations but incurred more overhead than the extrapolated start. Secant outer updates lowered intermediate misfits without reducing the number of passes at the target tolerances, while inter-pass warm-start adjustments produced little additional benefit. These negative results illustrate how measured performance, rather than plausible algorithmic improvements alone, determined which agent-assisted changes were retained.

\conclusions

The longstanding $O(N^2)$ scaling bottleneck in normal-mode GIA modeling is not an intrinsic physical limitation, but an artifact of historical load summation. By reformulating viscoelastic memory as an exact recursive state-space update, SLEvolve v1.0 reproduces the full-history solution to machine precision (differences below $10^{-11}$\,m) while reducing asymptotic complexity to $O(N)$.

This formulation-level breakthrough yields an immediate 20.6-fold acceleration in forward projections, which compounds to over three orders of magnitude (up to 4,000-fold) when combined with fewer, table-free and multithreaded spherical-harmonic transforms, and physics-inspired iteration starts. By compressing multi-millennial, multi-cycle simulations from days to seconds on a laptop, SLEvolve transforms gravitationally self-consistent sea-level modeling from an expensive specialized calculation into a practical tool for coupled ice-sheet modeling, large ensembles, and parameter inference, with potential applications to future differentiable or adjoint methods.

Finally, SLEvolve provides a case study in scientific software modernization: human-derived mathematical insight is combined with benchmark-harnessed AI coding agents to implement, test and optimize a numerically accurate solver.

\codedataavailability{SLEvolve v1.0.0 (Fortran and Python) is archived on Zenodo \citep{bao2026slevolve} under the GPL-3.0-or-later license; other versions, together with the manual, are available from the development repository at https://github.com/xiyuanbao/slevolve.
The Fortran solver is derived from the code of \citet{han2022}.
The ice forcing of the Antarctic case is from \citet{yousefi2022data}; the ICE-NADI-1 history is from \citet{lucas2026data}; and the archived output of \citet{han2022} and the ice history of Case 2 are from their OSF repository \citep{han2021osf}.} 

\authorcontribution{XB designed the study, derived the recursive load history formulation and the associated algorithms, developed the software with AI coding agents, performed the verification and wrote the paper. All authors discussed the results and reviewed and edited the manuscript.}

\competinginterests{The authors declare no competing interests.}

\begin{acknowledgements}
We thank the many GIA modelers who have made their 1-D sea-level codes publicly available (including Giorgio Spada, Jacqueline Austermann, Samuel Kachuck, and Holly Han), or by request (Sam Goldberg). In recent years a collaborative spirit has flourished within this community and we introduce SLEvolve v1.0 in this spirit. The SLEvolve v1.0 code was implemented, optimized and tested with AI coding agents (Claude Code with Claude Fable 5.1 and Claude Opus 5.5, Anthropic) under the direction of the first author; the extent of this use is described in Sect.~\ref{sec:ai}. The authors verified all results and take full responsibility for the code and the manuscript.
The project was supported by the Amazon Web Services (AWS) Impact Computing Project at the Harvard Data Science Initiative (HDSI) to X.B., S.M.M. and J.X.M. X.B. was also supported by Harvard Reginald A. Daly Postdoctoral Fellowship and Gordon and Betty Moore Foundation Postdoctoral Fellowship.
\end{acknowledgements}

\bibliographystyle{copernicus}
\bibliography{slevolve}

\end{document}